# The occultation of Arcturus in the Vatican

Costantino Sigismondi,
ICRANet and Observatorio Nacional RJ   sigismondi@icra.it

**Abstract**

The dome of Saint Peter's Basilica plays the role of the Moon during a stellar occultation and Arcturus is the target star. This occultation-like phenomenon is useful for introducing to occultation astronomy a class of student up to university level. It can be organized very easily at the convenience of the audience.
Techical and didactical aspects are discussed; the video is available at
http://www.youtube.com/watch?v=hIfsj7t-u-c and has been realized with an ordinary camcorder.

**Introduction**

Arcturus is not in the Zodiacal Catalogue and a lunar occultation is impossible.
The possibility of a TNO or asteroidal occultation is remote, but the observation of an occultation with a special landmark can be easily organized for presenting to an astronomy class the concept of stellar occultations.

The observation of the occultation of Arcturus recorded with an ordinary camcorder in Saint Peter's Square in the Vatican is here presented for its multiple didactical issues.
The large diffusion of mobile telephone camera allows nowadays to permit this kind of experience to all students, being the limit only in their will of learning more and more about astronomy and astronomical observations.

**Scintillation and electronic noise**

The use of a small camera in the observations of stars, when the video option is used, implies to know the scintillation effects in more detail.
With objective lenses of 2 cm this phenomenon is rather big, as shown e.g. in the occasion of the Venus occultation of 1 december 2008. [1]
The scintillation effect can be considered as the effect of the variation of the number of photons arriving at the objective. Their average value for unit time is N, the Poisson statistics implies a variation of $+/-\sqrt{(N)}$ during the same unit time interval.
If N is relatively small this variation is percentually significant.
The electronic noise also produces a similar effect,[2] and the combination of the two is the actual appearance of the star in the different frames of the video.
According to calculation for a zero magnitude star like Arcturus and a 2 cm diameter objective, [see e.g. Ref. 3] as the one of the SANYO CG9 camcorder, the Poissonian variations of detected luminosity of this star are much less important than the ones observed.
So we can deduce that the main responsible is the elecronic noise.

**Parallax effect**

The Moon, for the effect that it produces for the stellar occultation, can be considered at infinite distance with respect to our amateur telescopes diameters.
The occultation is nearly instantaneous, excepted for the stellar diameter effect.

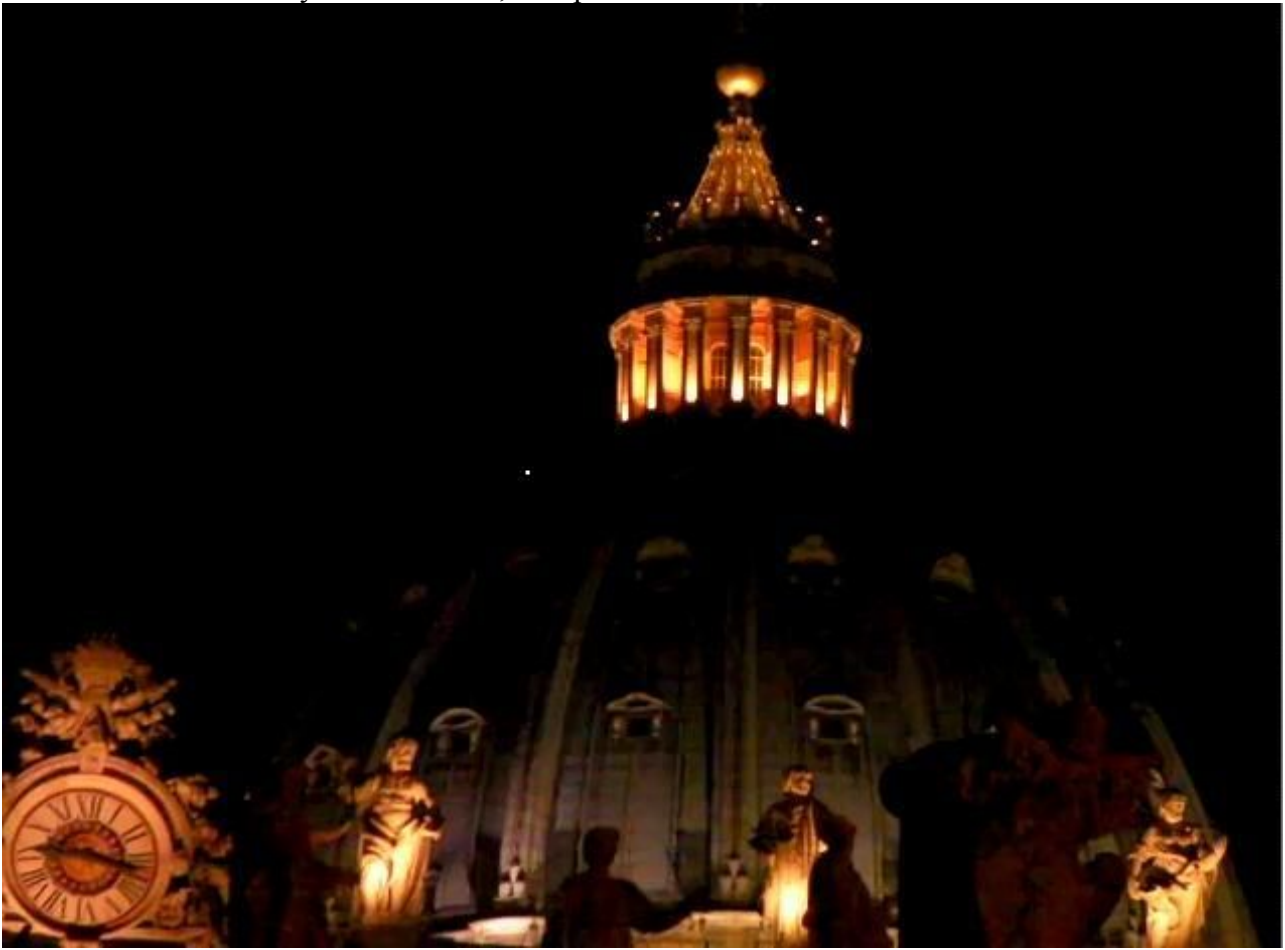

Fig. 1 The star approaching the Dome of Michelangelo, at 21:17 of 14 september 2013 as seen from behind the left fountain of St. Peter's square.
The idea of this measurement came on the evening of prevoius 7 of september while I was participating to the universal prayer for peace convocated by pope Francis in the St. Peter square, because from my position that occultation was visible.

When the occultation occurs with a finite distance target the parallax effect can be calculated from the angle formed by the objective at the distance with the target.
The procedure is described in the case of a similar Venus occultation at the Locarno Film Festival inauguration on
http://www.astroticino.ch/fileadmin/groups/astroticino/documenti/Meridiana_209_bassa.pdf . [4]
With a 2 cm camera, at about 400 m from the dome of St. Peter's Basilica we have a parallax effect of 1/20000 radians or 10 arsec, which correspond to 0.7 s of disappearance time of the star from the detector.
In this time interval, progressively, the light of the star decreases, while it goes behind the dome as seen from the various point of the objective lens.
We can easily imagine the profile of the dome moving in the opposite direction of the star, which determines the instant of disparition on each point of the lens.
This effect produces a progressive diminution of the star light, combined with the electronic noise and scintillation, and this diminution starts 0.7 s before the complete occultation.

As an example, for a 40 m telescope as the forthcoming ELT, we would expect an effect proportional to $40/400*10^6$ which is about $10^{-7}$ radians or 0.02 arcsec, which in the case of a lunar occultation it would last 0.04s = 1/25 s.

**Stellar diameter effect**

In the case of Arcturus the angular diameter is 0.02 arcsec,[5] i.e. a duration of the disparition of the stellar light of 0.02/15 s or 1.3 ms, well below the 1/30 s of maximum duration of a single frame, which is the video sampling time.
In the case of the Moon the relative velocity with the stars is about 0.5 arcsec/s, while in the case of landmarks the speed is the one of the daily motion: 15 arcsec/s.
So 0.02 arcsec would be covered in 0.04 s, i.e. 1/25 s, and visibile with an ordinary commercial camcorder.

The SANYO CG9 camera used for the video gets up to 1/60s of frame rate. Another model gets 1/300 s enough to appreciate this effect, there are other camera like CASIO Exilim capable of 1/1200 s on limitate period of time (10 s) and the possibility to film a stellar diameter becomes interesting.
Using photomultipliers the acquisition time is much faster, and this is the traditional way used for measuring stellar diameters.

**Video analysis**

It is possible to apply the same softwares of video analysis used in occultation astronomy to verify the parallactic effect.[1].
For a rapid analysis we can use simply Quicktime 7 which allows to inspect the single photogram.
In this case the last 10 photograms before the disparition are show the parallactic effect at 30fps it corresponds to 0.33 s, half of the expected value.

**Conclusions**

The didactic interest of this experiment is in the possibility to adapt it to other landmarks and other celestial objects, namely bright stars and planets, where their diameter's effect is much bigger, and its deconvolution from the parallactic effect becomes an interesting application of signal theory.
The observation can be easily organized at convenience of the audience.
This kind of angular measurement of planets is also very much stimulating for introducing students in positional astronomy and astrometry issues.

**Acknowledgements** To Orazio Converso for the edition in youtube